# lit-tag: a Shiny app for adding custom tags and notes to a citation database

Paul McElhany[1*], Kalina Grabb[2], and Madison M. Wood[3]

[1] Conservation Biology Division, Northwest Fisheries Science Center, National Marine Fisheries Service, National Oceanic and Atmospheric Administration, Seattle, Washington, United States of America

[2] Dalhousie University, Department of Oceanography, Halifax, Nova Scotia, Canada

[3] NOAA Ocean Acidification Program, Silver Spring, Maryland, United States of America

* Corresponding author

E-mail: paul.mcelhany@noaa.gov

# Abstract

To facilitate the review, evaluation and analysis of scientific literature, the lit-tag R Shiny application provides a convenient interface for users to generate a citation database with custom, user-defined tags and notes. Lit-tag is not subject-specific and is useful for any field of research. Starting with a table of citations exported from a Zotero library and a user-generated Excel file describing a set of tags and notes fields, lit-tag provides tools for assigning tags and notes to papers ("lit-tag-builder" module) and for exporting, graphing, and generating reports from the resulting database ("lit-tag-viewer" module). The app fills a need not met by the limited tagging tools available in bibliographic software and does not require database programming skills.

# Introduction

Scientific literature review and meta-analysis projects often involve summarizing the contents of many, often hundreds, of papers [1]. During the review, data are collected on many different attributes of each study (e.g., experiment type, treatment conditions, location, results, etc.). Although literature review projects may start out collecting this information in a simple spreadsheet, the approach quickly becomes unwieldy as the number of papers and attributes increase. A custom database with a user defined interface for data entry could be developed with general purpose database tools (e.g., [2,3]), but these often require programming skills and a substantial learning curve [4]. Dedicated citation tools, such as Zotero or Endnote, are designed to work with reference data, but have very limited tag and notes capability and are therefore insufficient for projects requiring a complex hierarchical tag structure with managed input options [5,6]. Artificial intelligence (AI) tools are also being developed (e.g. Elicit), however, 2025-era AI requires more oversight and expertise than manual methods [7,8].

At the same time researchers are compiling data on the contents of papers, they need to conveniently collect and use the full citation information for each paper. Zotero is an open-source reference management software which has tools for easily downloading citation information from the internet and for adding reference citations and bibliographies to documents [5]. Lit-tag combines the citation management strengths of Zotero with flexible tagging capabilities not currently available in the Zotero software. The lit-tag app uses a convenient user interface to link the contents of a reference library generated with Zotero to a set of user-defined paper attributes and notes stored in a CSV file that can be searched and visualized with tools in lit-tag or imported into another analysis platform like R or Python.

The lit-tag app was developed in R Shiny [9,10]. Source code is publicly available (https://github.com/pmcelhany/lit-tag) and an open-source online version of the app is hosted by NOAA Fisheries (https://connect.fisheries.noaa.gov/lit-tag/). A database generated and maintained by lit-tag consists of a simple CSV file with one row per paper and columns containing the citation information extracted from Zotero and all of the tags and notes generated by the user. In contrast to general purpose relational database tools like MySQL or Oracle, lit-tag requires no programming. The inputs are a Zotero export file and tags defined in a simple Excel file. Lit-tag's browser-based user interface provides a straightforward way to assign tag information to each paper and view results.

# Software design and work flow

The lit-tag app has two modules: 1) lit-tag-builder for generating, editing and updating the database and 2) lit-tag-viewer for generating output tables, graphs and reports from the database (Fig 1). The workflow for creating a new lit-tag database requires first building a library of references in Zotero [5]. Zotero was selected as the foundational bibliographic software for lit-tag because it is open source, has tools to easily collect citation information from the web and has plugins for inserting in-text citations and bibliographies in both Word and Google Docs. After creation of a Zotero library (or a collection within a library), citations are exported from Zotero as a CSV file, which contains both the citation information and unique keys generated by Zotero for every reference. The Zotero library or collection can be on any topic — the app is not restricted to any particular field of study.  In addition to the exported Zotero file, creation of a lit-tag database requires a user-generated Excel "categories" file, which defines the tags, tag options and notes fields that will be used to describe each of the papers in the database. The tags are defined by the user to collect information that is relevant to their particular research topic and study question — there are no pre-defined tags and there are no limits on the number of tabs and tags in the categories file. In the categories file, each tab of the Excel file defines a group of related tags. For example, a tab named "Location" might include tags for geographic region, elevation, and political jurisdiction. On each tab sheet, the first row contains the tag names, the second row describes the type of tag (e.g., single selection, multi-selection, date, or text field). For single or multi selection tags, rows three and greater contain the list of tag options. Restricting tag entries to a list of discrete tag options allows for categorical analysis of the information contained in the papers. The categories file also defines the "notes" fields, which are for free-form text information. Example Zotero CSV files and categories Excel files are available in the supplemental information.

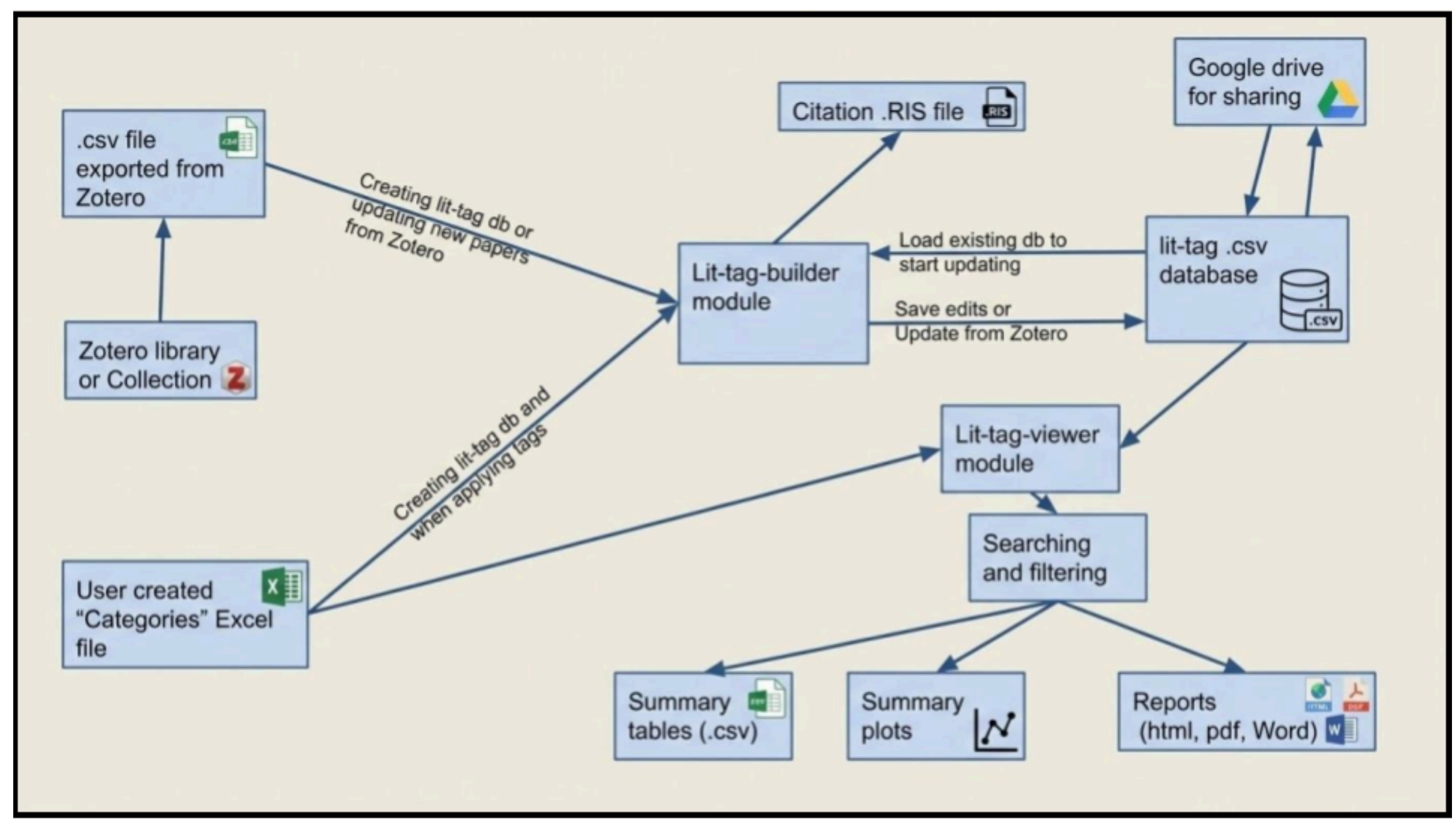

**Fig 1. Relationship between Zotero, lit-tag-builder module, lit-tag-viewer module and imported/exported files.** The lit-tag database is stored as a CSV file on a local computer, so it is not designed for simultaneous edits by multiple users. Sequential editing by multiple users can be accomplished by sharing the database on a platform such as Google Drive.

To create a new lit-tag database, the Zotero export file and categories file are uploaded on the "New database" tab of the lit-tag-builder module (Fig 2). To begin tagging, the newly created lit-tag database and categories file are uploaded on the "Tag edit" tab of the lit-tag-builder module (Fig 3). The Tag edit tab contains panels for selecting papers, viewing paper details and notes, and assigning tags to papers. Other tabs in the builder module have tools for 1) syncing with the Zotero database when new papers are added or papers are removed from the reference library, 2) database maintenance, 3) creating and linking to a new Zotero database , and 4) viewing the module user guide. The database maintenance tab includes options for viewing database contents (tag option counts), comparing database versions to explore the result of Zotero sync actions, deleting or replacing information in the database (Fig 4), and tools for combining databases for workflow scenarios in which multiple people are each tagging a subset of references as part of a larger project. The help tab includes a user's guide and example files.

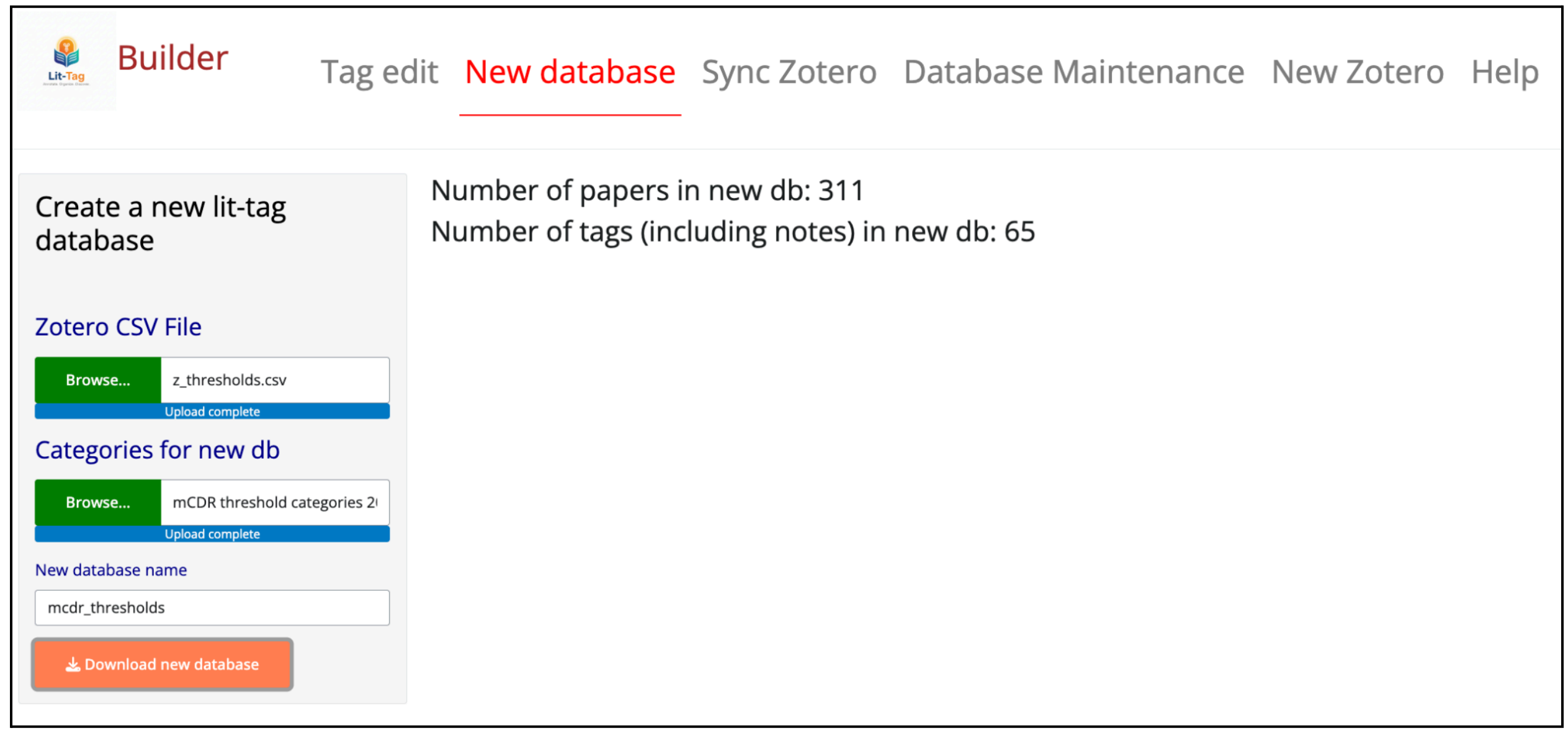

**Fig 2. Example screen shot of the "New database" tab of lit-tag-builder module user interface.**

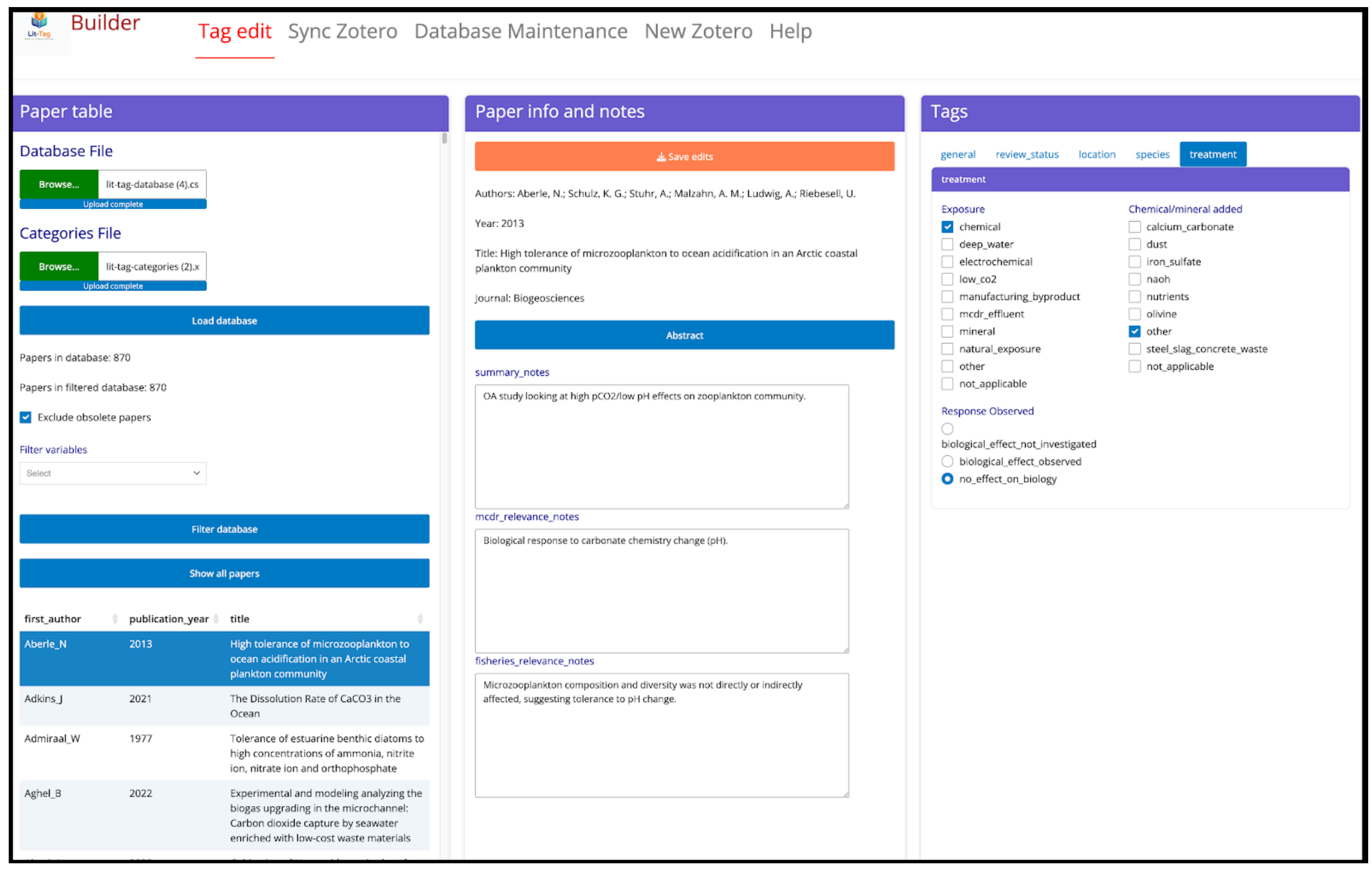

**Fig 3. Example screen shot of the "Tag edit" tab of lit-tag-builder module user interface.**

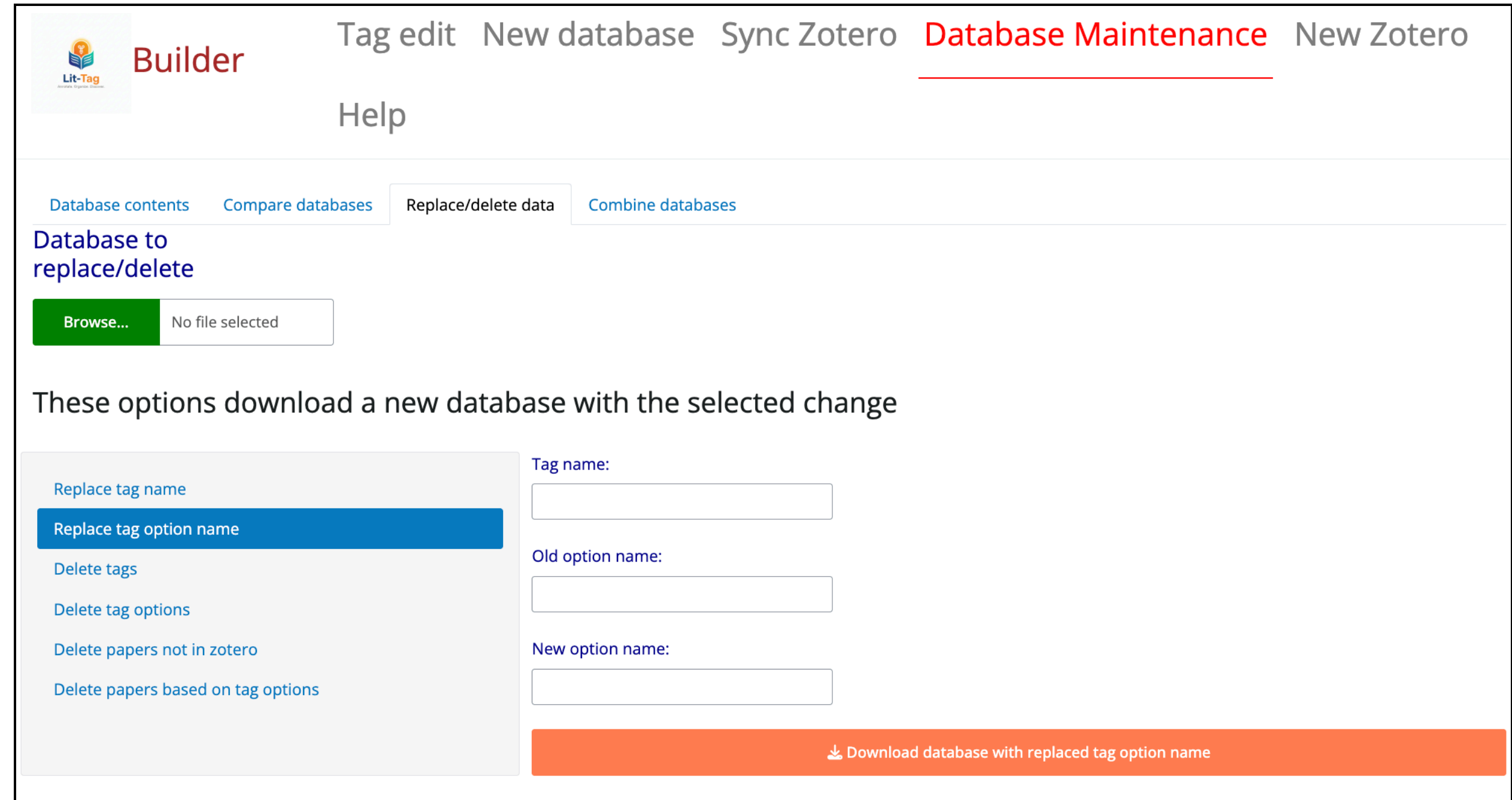

**Fig 4. Example screen shot of the “Replace tag option name” sub-tab for database maintenance in**

**the lit-tag-builder module user interface.**

Once a database has been tagged using the lit-tag-builder module, the lit-tag-viewer module contains
options for searching and filtering the database (including custom searches using R syntax), plotting
summary tables using any two tag variables for full and filtered datasets (Fig 5), and generating custom
tables (csv files) and reports (html, pdf or doc). The reports are generated using Quarto [11] with
user-selected content. The viewer module has convenient tools for exploring the database, but because the
database itself is a simple CSV file, analysis is not restricted to options in the viewer module. The CSV
file can be imported to R, Python or other analysis platforms. A comprehensive description of the
functionality of the builder and viewer modules is available in the app Help tabs, which include example
Zotero, categories and lit-tag database files.

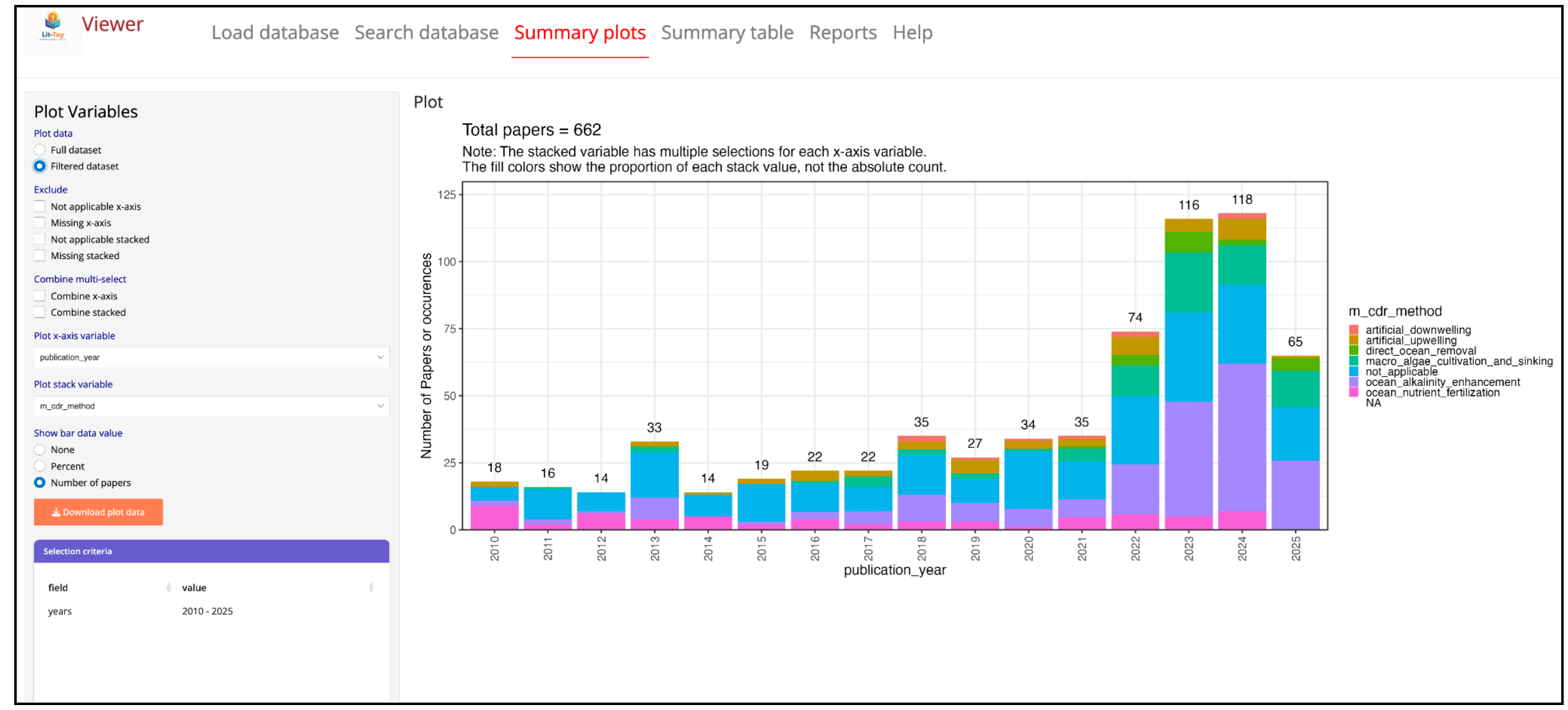

**Fig 5. Example screen shot of the “Summary plots” tab of the lit-tag-viewer module.**

# Example projects

To date, the lit-tag app has been used to develop two databases related to marine carbon dioxide removal (mCDR) research. mCDR technologies are currently being explored as an approach to extracting $CO_2$ from the atmosphere for climate change mitigation [12]. A lit-tag database was developed to compile information on papers related to the potential interaction of mCDR, fisheries and aquaculture [8]. The database ("mCDRxFisheries Literature Database") includes data on 870 references with 24 tags and 3 notes fields. For constrained selection tags, the number of options per tag ranged between 3 and 24. The database contained one date tag and two unconstrained text tags. A copy of the database embedded in a version of the lit-tag-viewer is available online (https://connect.fisheries.noaa.gov/mcdrxfisheries-lit-db-viewer/). The database supported a synthesis of the state of knowledge of the interaction of mCDR and fisheries [13].

A second mCDR study using lit-tag focused on biological thresholds for marine species exposed to the high pH, low $CO_2$ or high alkalinity seawater conditions that can be created by some mCDR methods [14]. This database ("mCDR Thresholds Literature Database") contained 310 references, 62 tags and 5 notes fields. The database was used to compile quantitative information extracted from each of the studies on species tolerance to altered seawater chemistry.

# Limitations

The lit-tag database is stored on a local drive, which creates challenges when multiple people are collaborating on a single tagging project (i.e. the app does not support simultaneous edits of a single database). If only a few people are working on a tagging project, as was the case with the mCDRxFisheries Literature Database, sequential editing is possible, with the latest version of the database uploaded to to a shared repository (e.g. Google Drive), then downloaded for the next round of editing. To aid in keeping track of the database version, a UTC timestamp is appended to the file name
with every save of the database in the lit-tag-builder module. An alternative approach for multiple taggers
was used to create the mCDR Thresholds Literature Database. The Zotero library was divided into
sub-collections and each tagger created a separate database for their sub-collection. In the case of the
mCDR Thresholds project, the sub-collections were based on species taxonomic groups, but the
sub-collections could be based on anything, including a simple division of workload. For comprehensive
project analysis of the entire library, the separate databases can be easily combined using the tools in the
lit-tag-builder module.

In addition to the constraint on simultaneous edits, a lit-tag database is restricted to linking to a single
Zotero library. Lit-tag uses the unique keys generated by Zotero for linking the database to Zotero for
syncing operations. Zotero creates a unique key every time a reference is added to a library, which means
that different libraries will have different unique keys, even for the exact same citation. The easiest way to
address this library-specificity is by using an online shared Zotero library that is accessible to all taggers.
If it is necessary to link an existing lit-tag database to a new Zotero library, the lit-tag-builder module has
tools for creating a new Zotero library and linking its keys to a copy of the original lit-tag database.

# Discussion and conclusion

The lit-tag app provides a flexible, easy to use tool for tagging papers as part of a literature review or
meta-analysis. The lit-tag workflow integrates the robust tools of a fully developed bibliographic software
platform for reference acquisition and within-document citation with a customized user interface for
tagging and recording notes. Lit-tag provides for the interactive workflow that happens with real-word
projects. Papers can be easily added or removed from the analysis and it is possible to change what data
are collected on each of the papers as the project evolves. For example, during creation of the databases
for the mCDR projects, we often found the need to add more tags to collect additional information on the papers or to change the section options for a particular tag as we refine tag definitions, or occasionally, we deleted tags that turned out to not provide any relevant information. Because the categories file defining tags is independent of the database file, these sorts of modifications were transparent, archived, and easy to execute. Ultimately, lit-tag offers a powerful, accessible, and subject-agnostic solution for researchers needing to perform data-intensive literature synthesis.

# Acknowledgements

We would like to thank the International Council for the Exploration of the Sea (ICES) mCDR x Fisheries Workshop group for inspiration and app testing.